\documentclass[doublecol,linenumbers]{epl2} 

\usepackage{graphicx,epstopdf}
\usepackage{amsmath}
\usepackage{amssymb}
\usepackage{mathtools}
\usepackage{braket}
\usepackage{color}
\usepackage{float}
\usepackage{epstopdf}
\usepackage{amsthm}
\usepackage{bm}
\usepackage{graphics,epstopdf}
\usepackage{color}
\usepackage[usenames,dvipsnames,svgnames]{pstricks}
\usepackage{epsfig}
\usepackage{pst-grad} 
\usepackage{pst-plot}

\title{Pancharatnam Phase Deficit can detect Macroscopic Entanglement}

\author{Namrata Shukla and Arun Kumar Pati}
\shortauthor{F. Author \etal}

\institute{ Harish-Chandra Research Institute\\
Chhatnag Road, Jhunsi, Allahabad 211
019, India
}

\abstract{The Pancharatnam phase deficit is defined as the difference between the Pancharatnam phase acquired by 
the global system and sum of the Pancharatnam phases acquired by subsystems during local unitary evolutions. We show that a non-zero value of the Pancharatnam phase deficit for a composite quantum system can be a 
signature of quantum entanglement. In the context of macroscopic quantum systems, we illustrate how the Pancharatnam phase deficit 
can be used to detect macroscopically entangled states. In particular, we use the Pancharatnam phase deficit to detect the entanglement for macroscopic superposition of coherent states. Furthermore, we 
show that by measuring the Pancharatnam phase deficit one can measure the concurrence of two spin singlets between distant boundaries.}

\begin{document}

\maketitle

\section{Introduction}

Berry's phase, alternatively called as the Panchratnam phase \cite{Berryoriginal} has been a subject of great importance in quantum mechanics. 
The Barry phase has been generalized for non adiabatic but cyclic
evolutions of quantum system \cite{Aharnov}
and for noncyclic evolutions \cite{Samuel-Bhandari}. In general, without taking into account the initial ideas of adiabaticity, cyclicity, 
and unitarity for geometric phase, if a quantal system undergoes an evolution, we can can compute
the phase difference 
in the process of evolution by the inner product between the states before and after
evolution, with an exception of initial and final states being orthogonal \cite{Panchrantnam,Simon}.
The geometric phase has been studied for various quantum 
systems \cite{Tong,Yi, Barry1} and it has important 
applications \cite{ArunPati,Jones,Ekert,Sjoqvist,Jain, Pachos1,Pachos2,Carollo,ArunPati2,ArunPati21} in literature.

Quantum entanglement \cite {Cerf, Horodecki} being at the heart of quantum
information theory, its detection plays very important role in quantum systems. 
The connection of geometric phase and entanglement has also been studied before \cite {Calvani, Hartley, Basu}
and the noncyclic geometric phase for entangled 
states was calculated by Sj\"oqvist \cite{Sjoqvist1}. Since, in principle, quantum mechanics may permit the existence of a macroscopic object which is 
in a quantum superposition, it has become very important to study the macroscopic 
quantumness and macroscopic entanglement. Macroscopic quantumness is one of the crucial aspects of the present research on quantum systems and it is strongly connected with the quantum to 
classical transition and the measurement problem \cite{Calvani}. There have been several proposals to quantify macroscopicity based on
different criteria involving effective number of particles in the superposition states, dishtinguishability and operational interpretations
\cite{Leggett, Dur-Cirac, Shimizu, Bjork}. 
In recent years, macroscopic quantum states have been proposed in various systems such as superconductors \cite{Rouse,Clarke,Silvestrini,Nakamura}, 
nanoscale qubits \cite{Friedman,Wernsdorfer}, trapped ions \cite{Monroe}, and photonic qubits in microwave cavity \cite{Brune}.
Constructing such macroscopically superposed states \cite{Liu, Bayat}, studying and detecting entanglement in such macroscopic states are need of the hour
to understand the extent of applicability of quantum mechanics.

In this letter we make use of the notion of the Pancharatnam phase deficit 
during arbitrary local unitary evolution of a macroscopic quantum system and show how one can detect entanglement using the Pancharatnam phase deficit. 
A bipartite pure quantum system, subjected to local unitary evolution acquires a Pancharatnam phase during the
evolution. However, the individual subsystems also acquire 
their respective Pancharatnam phases during the evolution. We show that if the state is
separable, then the difference in the global Pancharatnam 
phase and sum of the local Pancharatnam phases, which we call as the Pancharatnam phase
deficit, will be zero. Therefore, a nonzero value of the 
Pancharatnam phase deficit acquired by the pure composite system is a signature of
entanglement. We show how the Pancharatnam phase deficit can be a signature of entanglement in macroscopic superposition of 
quantum states. Interestingly, in some specific examples, we show that the entropy of entanglement and concurrence for macroscopic system can be expressed
solely using the Pancharatnam phase deficit. \\

\section{Pancharatnam Phase Deficit for Bipartite States}

Consider an arbitrary initial product state $|\Psi(0) \rangle_{AB} =
|\psi(0) \rangle_A \otimes |\phi(0) \rangle_B \in {\cal H}_A \otimes {\cal H}_B$ of a bipartite composite system. Let this
evolve under local unitary evolution, namely, 
 the system is subject to Hamiltonian of the form $H = H_A \otimes I + I \otimes H_B$. 
Under the action of the  Hamiltonian the state evolves unitarily as $|\Psi(0)
\rangle_{AB} \rightarrow 
|\Psi(t) \rangle_{AB} = U(t) \ket{\psi(0)}_A \otimes  V(t) \ket{\phi(0)}_B$, where $U(t)=e^{-iH_At/\hbar}$
and $V(t)=e^{-iH_Bt/\hbar}$.
In this section we consider the local unitary evolution of a bipartite state and try
to characterize 
the entanglement in terms of the difference of the Pancharatnam phases acquired by the
system in global and local scenarios. The Pancharatnam phase acquired by the composite system is given by 
\begin{eqnarray}
\Phi_T^{AB}&=&Arg( \langle \Psi(0)|\Psi(t)\rangle).
\end{eqnarray}
Now, the Pancharatnam phases acquired by the  subsystem $A$ and $B$, respectively, are given by 
\begin{equation}
\Phi_T^A = Arg( \langle \psi(0)|\psi(t)\rangle),~\Phi_T^B = Arg( \langle \phi(0)|\phi(t)\rangle).
\end{equation}

The Pancharatnam phase deficit is defined as the difference of the Pancharatnam phase
acquired by the composite system and the sum of the 
Pancharatnam phases acquired by the local subsystems, i.e., 
\begin{eqnarray}
\Delta&=&\Phi_T^{AB}-[\Phi_T^A+\Phi_T^B].
\label{difference}
\end{eqnarray}
For product states, the Pancharatnam phase deficit is identically  zero. Therefore, if
it is nonzero, then the state will be entangled. Hence, a non-zero 
Pancharatnam phase deficit is shown to be a signature of entanglement in 
the composite quantum system. \\

Suppose we have a general bipartite state in $\cal H_A^\text{\it N} \otimes \cal H_B^\text{\it M}$ for the
composite system. Using the Schmidt decomposition 
theorem we can write the combined bipartite state (with $N<M$) as
\begin{equation}
|\Psi  \rangle_{AB} =  \sum_{n=1}^N \sqrt{\lambda_n } |a_n\rangle |b_n \rangle,
\end{equation}
where $\lambda_n$'s are the Schmidt coefficients with 
$\sum_n \lambda_n =1$, $|a_n \rangle \in {\cal H}_A$ and 
$ |b_n \rangle \in {\cal H}_B$ are the orthonormal Schmidt basis. Let the state evolve as $
|\Psi\rangle_{AB}\rightarrow U(t) \otimes V(t) |\Psi\rangle_{AB}$. 
Under this local unitary evolution, 
the Pancharatnam phase for the composite state is given by 
\begin{eqnarray}
\Phi_T^{AB} &=& Arg( \langle \Psi|U(t) \otimes V(t) \Psi \rangle).
\end{eqnarray}
To find the Pancharatnam phases for subsystems, we have to recourse to the notion of the 
mixed state Pancharatnam phase \cite{ArunPati21}.
For the subsystems $A$ and $B$ this is given by 
\begin{equation}
 \Phi_T^A =Arg (Tr[\rho_A U(t)]),~\Phi_T^B = Arg (Tr[\rho_B V(t)]).
\end{equation}
Therefore, for a general bipartite entangled state the Pancharatnam phase deficit is
given by
\begin{eqnarray}
\Delta &=&\Phi_T^{AB}-[\Phi_T^A+\Phi_T^B]\nonumber\\ 
&=&\tan^{-1}\big[ \frac{\sum_{kl} \sqrt{\lambda_k \lambda_l} ( Im U_{kl} Re
V_{kl} + Re U_{kl} Im V_{kl} )}
{\sum_{kl} \sqrt{\lambda_k \lambda_l} (Re U_{kl} Re V_{kl} - Im U_{kl} Im V_{kl}) }
\big] \nonumber\\
&&- \tan^{-1}\big[ \frac{\sum_{k}  \lambda_k Im U_{kk} }{\sum_{k} \lambda_k  Re
U_{kk} } \big] -
\tan^{-1}\big[ \frac{\sum_{k} \lambda_k  Im V_{kk} }{\sum_{k} \lambda_k  Re V_{kk} }
\big],\nonumber\\
\label{difference_bipartite}
\end{eqnarray}
where $U_{kl}=\langle a_k|U|a_l\rangle,$ and $ V_{kl}=\langle b_k|V|b_l\rangle$.
Note that for entangled state under local unitary evolutions, the global dynamical
phase is equal to the sum of 
local dynamical phases for the subsystems $A$ and $B$, i.e., $\Phi_D^{AB} = \Phi_D^A
+ \Phi_D^B$.
Hence, if one defines the dynamical phase deficit that will 
be always zero for entangled as well as separable states. Therefore, dynamical phase
deficit cannot detect entanglement of bipartite
states. It is the Pancharatnam phase that captures the notion of
entangled state via a non-zero Pancharatnam phase deficit. 


In the present method, we can detect entanglement without state tomography or measuring the spin observables. 
Also our method can detect entanglement and measure the entropy of entanglement 
for two spin half particles using the Pancharatnam phase deficit. Since we do not need cyclic or adiabatic conditions 
for the Pancharatnam phase deficit, our proposal is more general. Our result can be applied to detect entanglement across microscopic-microscopic
as well as microscopic-macroscopic system partitions. For example, one can imagine that the subsystem $A$ is made of large number of elementary sub-sub systems 
(say $N\sim10^{4}$) and the multiparticle states are orthonormal and have macroscopically distinct average value of a physical quantity. Such systems can also be imagined 
with quantum simulators \cite{Buluta} which will be able to emulate a large number of quantum systems and can exhibit quantum properties being a quantum system itself.

\section{Pancharatnam Phase Deficit for Multi-particles}
Suppose, we have a multi-particle 
state which is not in a product state. Then, let it undergo local unitary evolutions,
i.e., $\ket{\Psi}_{A_1A_2....A_N} \rightarrow 
U_1 \otimes U_2 \otimes \cdots U_N  \ket{\Psi}_{A_1A_2....A_N}$. The Pancharatnam phase
acquired by the composite state is given by 
\begin{equation}
\Phi_T^{A_i..A_N}=Arg( \langle \Psi |U_1 \otimes U_2 \otimes \cdots U_N  \ket{\Psi}_{A_1, A_2....A_N},
\end{equation}
However, the constituent systems would also acquire their respective Pancharatnam
phases during local evolutions and we denote them
by $\Phi_T^{A_i}$, where $A_i=A_1, A_2, \ldots A_N$. The local Pancharatnam phase is given by 
\begin{align}
\Phi_T^{A_i} = Arg( Tr[\rho_{A_i} U_i] ),
\end{align}
where $i=1,2, \ldots N$. 
The Pancharatnam phase deficit for $N$ particle is given by 
\begin{eqnarray}
\Delta&=&\Phi_T^{A_i..A_N}-\bigg[\sum_i \Phi_T^{A_i}\bigg] \nonumber\\
&=& Arg (\bra{\Psi}|U_1 \otimes U_2 \otimes \cdots U_N  \ket{\Psi}_{A_1, A_2....A_N})\nonumber \\
&&-\sum_i Arg( Tr[\rho_{A_i} U_i] )
\end{eqnarray}

If $N$-particle state is a pure product state, then $ \Delta=0$. This implies that
a non-zero $\Delta$ signifies that the state
$\Psi_{A_1A_2 \ldots A_N}$ is entangled. \\

Our method can be applied to detect entangled states in quantum interferometry. In Mach-Zehnder interferometer set up, 
we can measure the relative phase acquired during the local unitary evolution 
by the global system and the local subsystems. Relative phase being easier to measure in interferometry, this is a convenient 
method to detect entanglement in composite systems. Whether our method of detecting entanglement using the Pancharatnam 
phase deficit is robust for decoherence, will be explored in future.
It should be stressed that the explorations in macroscopic quantum systems
is strongly related with the quantum to classical transition and the
measurement problem. For example, it has been shown the existence of the
environment which usually causes decoherence, can induce the Berry phase
and entanglement between the system and the environment can be measured
with the help of the Barry phase \cite{Calvani1, Liuzzo}.

\section{Pancharatnam Phase Deficit for Entangled Macroscopic and Microscopic Systems}
Using the single photon quantum-injected optical parametric amplification, the entanglement between the microscopic qubit and the macroscopic 
part has been obtained experimentally \cite{Martini}. The macroscopic part consists of large number of photons. Below,
we show how our result can be applied to detect the micro-macro entangled state. Consider a system composed of macroscopic and microscopic parts in an entangled state \cite{Martini} such as, 
\begin{equation}
|\Psi\rangle_{AB}=\sqrt{\lambda_0}~|\psi\rangle\otimes|0\rangle+\sqrt{\lambda_1}~|\bar{\psi}\rangle\otimes|1\rangle, \label{wavevector_ex0}
\end{equation}
where the subsystem $A$ is a macroscopic quantum system (contains photon number of the order $10^4$) with $|\psi\rangle$, $|\bar{\psi}\rangle$ are the 
mutually orthogonal multiparticle macroscopic state (See for details \cite{Martini}) and the subsystem $B$ is a microscopic system. Let the system evolve under local time-independent Hamiltonian 
\begin{equation}
H=\epsilon_1 |\bar{\psi}\rangle\langle\bar{\psi}|\otimes I_B+\epsilon_2 I_A\otimes|1\rangle\langle1|,
\end{equation}
where $\epsilon_1, \epsilon_2>0$ and $I_A, I_B$ are the identity operators on $\cal H_A$ and $\cal H_B$. 
Under this Hamiltonian, the state $|\Psi\rangle_{AB}$ undergoes 
local unitary evolution given by $|\Psi\rangle_{AB}\rightarrow U\otimes V |\Psi\rangle$ and the evolved state is
\begin{eqnarray}
|\Psi\rangle_{AB}(t)&=&e^{-i\epsilon_1t/\hbar |\bar{\psi}\rangle\langle\bar{\psi}|}\otimes e^{-i\epsilon_2t/\hbar |1\rangle\langle1|}|\Psi\rangle_{AB} \nonumber \\
&=&\sqrt{\lambda_0}~|\psi\rangle\otimes|0\rangle+\sqrt{\lambda_1}~e^{-i(g_1+g_2)}|\bar{\psi}\rangle|1\rangle,\nonumber\\
\label{evolved_wavevector_ex0}
\end{eqnarray}
where $g_1=\epsilon_1 t/\hbar$ and $g_2=\epsilon_2 t/\hbar$. We have used the unitary evolution operators as
\begin{eqnarray}
U&=&e^{-i\epsilon_1 t/\hbar |\bar{\psi}\rangle\langle\bar{\psi}|}=(I-|\bar{\psi}\rangle\langle\bar{\psi}|)+e^{-i\epsilon_1 t/\hbar}|\bar{\psi}\rangle\langle\bar{\psi}|,\nonumber\\
V&=&e^{-i\epsilon_1 t/\hbar |1\rangle\langle1|}=(I-|1\rangle\langle1|)+e^{-i\epsilon_1 t/\hbar}|1\rangle\langle1|.
\label{unitary_ex0}
\end{eqnarray}
The Pancharatnam phase acquired by the composite system during this evolution turns out to be
\begin{equation}
\Phi_T^{AB}=tan^{-1}\bigg[\frac{-\lambda_1 \sin(\epsilon_1+\epsilon_2)t/\hbar}{\lambda_0+\lambda_1 \cos(\epsilon_1+\epsilon_2)t/\hbar} \bigg].
\label{globalphase_ex0}
\end{equation}
The local Pancharatnam phase shifts are given by
\begin{eqnarray}
\Phi_T^A&=&tan^{-1}\bigg[\frac{-\lambda_1 \sin \epsilon_1 t/\hbar}{1+\lambda_1(\cos \epsilon_1 t/\hbar-1)}\bigg],\\ \label{localphase1_ex0}
\Phi_T^B&=&tan^{-1}\bigg[\frac{-\lambda_1 \sin \epsilon_2 t/\hbar}{1+\lambda_1(\cos \epsilon t_2/\hbar-1)}\bigg].\label{localphase2_ex0}
\end{eqnarray}
Therefore, the Pancharatnam phase deficit is given by
\begin{eqnarray}
\Delta&=&\tan^{-1}\bigg[\frac{-\lambda_1 \sin(\epsilon_1+\epsilon_2)t/\hbar}{\lambda_0+\lambda_1 \cos(\epsilon_1+\epsilon_2)t/\hbar}\bigg]\nonumber \\
&&-\tan^{-1}\bigg[\frac{-\lambda_1 \sin \epsilon_1 t/\hbar}{1+\lambda_1(\cos \epsilon_1 t/\hbar-1)}\bigg]\nonumber\\
&&-\tan^{-1}\bigg[\frac{-\lambda_1 \sin \epsilon_2 t/\hbar}{1+\lambda_1(\cos \epsilon t_2/\hbar-1)}\bigg].\label{difference_ex0}
\end{eqnarray}
Now, if we choose the parameters in the Hamiltonian such that $\epsilon_1=\epsilon_2$ and $2\epsilon t/\hbar=\pi$, then we get
\begin{equation}
\lambda_0=\frac{1}{1+\tan\Delta/2},~\lambda_1=\frac{\tan\Delta/2}{1+\tan{\Delta/2}}.
\label{coefficients}
\end{equation}
The entropy of entanglement between the macroscopic and microscopic parts for this system is given by $E_N=-Tr(\rho_A \ln \rho_A)=-Tr(\rho_B \ln \rho_B)$.
On using Eqs. \eqref{coefficients} in this expression, we can express the entanglement entropy solely in 
terms of the Pancharatnam phase deficit. This is given by
\begin{eqnarray}
E_N&=&-\frac{1}{1+\tan\Delta/2}\ln \Big(\frac{1}{1+\tan\Delta/2}\Big)\nonumber\\
&&-\frac{\tan\Delta/2}{1+\tan\Delta/2} \ln \Big(\frac{\tan\Delta/2}{1+\tan\Delta/2}\Big).
\label{concurrence_PD}
\end{eqnarray}
Thus, by suitably choosing the parameters in the Hamiltonian,
one can directly measure the entanglement with the help of the Pancharatnam phase deficit in quantum interferometry.

\section{Pancharatnam Phase deficit for macroscopic superposition of coherent states}
Liu {\it et al.} \cite{Liu} have proposed the generation of superposition of
macroscopically distinguishable 
coherent states using a microwave cavity with a superconducting charge qubit. The
evolved entangled state at 
time $t$ generated can be written in the form
\begin{equation}
|\Psi(t)\rangle_{AB}=\frac{1}{\sqrt{2}}
\big(k|\alpha_-\rangle|g\rangle+k^*|\alpha_+\rangle|e\rangle\big),
\label{wavevector_ex2}
\end{equation}
where $ k=e^{-i \bm\psi}$ is a complex constant with $\bm\psi=$~Im$\Big[\frac{\zeta E_J}{\hbar \omega} \alpha'(1-e^{i\omega t})\Big]$ 
such that $ \Omega=\zeta^* E_J/\hbar$ is complex Rabi frequency (see for details \cite{Liu}). The injected coherent field 
$|\alpha'\rangle$ has real amplitude $\alpha'$ and $\omega$ is the frequency of single mode cavity field. $|g\rangle$ and $|e\rangle$ denote 
the ground state and the excited state of the atom, respectively. In order to test the entanglement in this given system, let this state evolve under $U\otimes V$ with $U$
and $V$ chosen as phase shifting operator and Pauli spin operator
$U= e^{-i\theta \hat N}$ and $V=\sigma_x$, respectively.
Here, $\hat N$ is number operator and $\sigma_x$ is the Pauli matrix.~The transition amplitude under local evolution of the
entangled state in Eq.\eqref{wavevector_ex2} is given by
\begin{eqnarray}
&&Tr[\rho_{AB}(t) U\otimes V]\nonumber\\
&&=\frac{1}{2} \exp \big(-1/2(|\alpha_-|^2+|\alpha_+|^2)\big) \nonumber\\
&&\Big[\Big(k^2 \exp\big(|\alpha_-||\alpha_+|\cos(\xi-\theta)\big)\cos\big(|\alpha_-||\alpha_+|\sin(\xi-\theta)\big)\nonumber\\
&&+k^{*2} \exp\big(|\alpha_-||\alpha_+|\cos(\xi+\theta)\big)\cos\big(|\alpha_-||\alpha_+|\sin(\xi+\theta)\big)\Big)\nonumber \\
&&+i\Big(k^2 \exp\big(|\alpha_-||\alpha_+|\cos(\xi-\theta)\big)\sin\big(|\alpha_-||\alpha_+|\sin(\xi-\theta)\big)\nonumber\\
&&-k^{*2} \exp\big(|\alpha_-||\alpha_+|\cos(\xi+\theta)\big)\sin\big(|\alpha_-||\alpha_+|\sin(\xi+\theta)\big)\Big)\Big],\nonumber\\
\label{tracerhoAB_ex2}
\end{eqnarray}
on using the notations $\xi=\chi_1-\chi_2$, and
$\alpha_-=|\alpha_-|e^{i\chi_1},~\alpha_+=|\alpha_+|e^{i\chi_2}$. 
The local phases acquired during this unitary evolution by the individual quantum
systems can be given by
\begin{eqnarray}
&&Tr[\rho_A(t)U]\nonumber\\
&&=\frac{1}{2}\Big[\Big(\exp (|\alpha_-|^2(\cos\theta-1))\cos(|\alpha_-|^2\sin\theta)\nonumber\\
&&+\exp(|\alpha_+|^2(\cos\theta-1)\cos(|\alpha_+|^2\sin\theta)\Big)\nonumber\\
&&-i\Big(\exp (|\alpha_-|^2(\cos\theta-1))\sin(|\alpha_-|^2\sin\theta)\nonumber\\
&&+\exp(|\alpha_+|^2(\cos\theta-1)\sin(|\alpha_+|^2\sin\theta)\Big)\Big],
\end{eqnarray}
\begin{eqnarray}
&&Tr[\rho_B(t)V]\nonumber\\
&&=\frac{1}{2}\exp\big(-1/2(|\alpha_-|^2+|\alpha_+|^2)+|\alpha_-||\alpha_+|\cos\xi\big)\nonumber\\
&&\Big[(k^2+k^{*2})\cos(|\alpha_-||\alpha_+|\sin\xi)\nonumber\\
&&+i(k^2-k^{*2})\sin(|\alpha_-||\alpha_+|\sin\xi)\Big].
\label{tracerhoB_ex2} 
\end{eqnarray}
Eq. \eqref{difference_bipartite} gives the Pancharatnam phase deficit for this entangled state after unitary evolution as
\begin{eqnarray}
&&tan^{-1}\left[
\frac{
\splitfrac{
\Big(k^2 \exp\big(n\cos(\xi-\theta)\big)\sin\big(n\sin(\xi-\theta)\big)
}{-k^{*2} \exp\big(n\cos(\xi+\theta)\big)\sin\big(n\sin(\xi+\theta)\big)\Big)
}
}{
\splitfrac{
\Big(k^2 \exp\big(n\cos(\xi-\theta)\big)\cos\big(n\sin(\xi-\theta)\big)
}{+k^{*2} \exp\big(n\cos(\xi+\theta)\big)\cos\big(n\sin(\xi+\theta)\big)\Big)
}
}
\right]\nonumber\\
&&-tan^{-1}\left[
\frac{
\splitfrac{
\Big(-\exp (n_-(\cos\theta-1))\sin(n_-\sin\theta)
}{-\exp(n_+(\cos\theta-1)\sin(n_+\sin\theta)\Big)
}
}{
\splitfrac{
\Big(\exp (n_-(\cos\theta-1))\cos(n_-\sin\theta)
}{+\exp(n_+(\cos\theta-1)\cos(n_+\sin\theta)\Big)
}
}
\right]\nonumber\\
&&-tan^{-1}\Big[\frac{(k^2+k^{*2})\sin(n\sin\xi)}{(k^2-k^{*2})\cos(n\sin\xi)}\Big],\nonumber
\end{eqnarray}
where $ n_-=|\alpha_-|^2, n_+=|\alpha_+|^2$ and $\sqrt{n_-n_+}=n$.
To compare with the other measures of entanglement, we calculate the entropy of entanglement for this state as
\begin{eqnarray}
E_N&=&-Tr(\rho_A \ln \rho_A)\nonumber\\
&=&\frac{1}{2 \ln 2} \Big(kk^*e^{\frac{-n_-}{2}\frac{-n_+}{2}+n_-n_+ \cos\xi}-1\Big)\nonumber\\
&&\ln \Big[\frac{1}{2}\Big(1-kk^*e^{\frac{-n_-}{2}\frac{-n_+}{2}+n_-n_+ \cos\xi}\Big)\Big] \nonumber\\
&&-\frac{1}{2 \ln 2} \Big(1+kk^*e^{\frac{-n_-}{2}\frac{-n_+}{2}+n_-n_+ \cos\xi}\Big)\nonumber\\
&&\ln \Big[\frac{1}{2}\Big(1+kk^*e^{\frac{-n_-}{2}\frac{-n_+}{2}+n_-n_+ \cos\xi}\Big)\Big]\nonumber\\
\label{entropy_ex2}
\end{eqnarray}

\begin{figure} \label{Plot_ex1}
\centering
\includegraphics[scale=0.9]{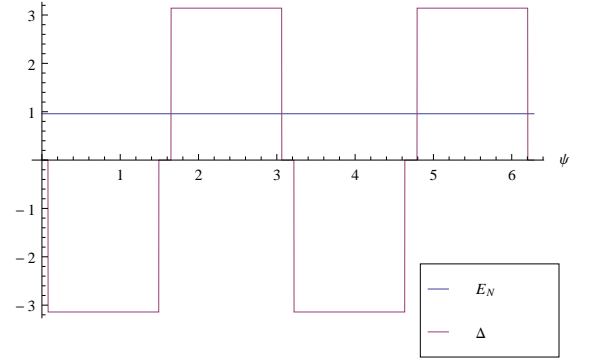}
\caption{(Color Online) Variation of the Pancharatnam phase deficit $\Delta$ and entropy of entanglement $E_N$ with $\bm\psi$ for $|\alpha_-|^2=2,|\alpha_+|^2=1,~\xi=\pi/4$ and
$\theta=\pi$.The blue line indicates the entropy of entanglement and the magenta line indicates the value of Pancharatnam phase deficit.} 
\end{figure}
Now, we plot the entropy of entanglement and the Pancharatnam phase deficit for some specific values of parameters and very small photon numbers
in the two coherent modes in Fig. 1. It is evident from the illustration that the non zero phase deficit is 
a sufficient condition for entanglement in a macroscopic quantum system.

\section{Pancharatnam Phase deficit for two distant boundary spins}

Bayat {\it et al.} \cite{Bayat} have proposed a method to create high entanglement between very distant boundary 
spins which is generated by suddenly connecting two long Kondo spin chains. As an example, they consider two spin singlets 
each formed by only two spins interacting with a Heisenberg interaction of strength $J_1$ and $J_2$, respectively. 
One may generate high entanglement between the boundary spins 1 and 4, by just turning on an interaction $J_m$ between 
the two spins 2 and 3. On choosing a specific condition $ J_m=J_1+J_2$ at certain time, the evolved state at time $t$ 
(up to a global phase) for boundary spins 1 and 4 of the system, is given by
\begin{eqnarray}
|\Psi(t)\rangle_{1234}=\frac{-i \sin (2 J_m t)}{2} \big(|0011\rangle+|1100\rangle\big)-\frac{\cos(2 J_m t)}{2}\nonumber \\
\times\big(|1001\rangle+|0110\rangle\big)+\frac{e^{i2J_mt}}{2}\big(|0101\rangle+|1010\rangle\big).\nonumber\\
\label{wavevector_ex4}
\end{eqnarray}
The concurrence $E$ \cite{Hill} for spins 1 and 4 is given by
\begin{equation}
E=\max\bigg\{0, \frac{1-3\cos(4J_mt)}{4}\bigg\}.
\label{concurrence}
\end{equation}
Using the method discussed in this paper, we would 
see that not only the entanglement can be detected by nonzero phase deficit but the maximum phase deficit 
reflects the maximum entanglement. For the state given by Eq. \eqref{wavevector_ex4} we choose the local
unitary evolution under the following Hamiltonian in order to connect the entanglement between the spins 1 and 4 and the Pancharatnam phase deficit. 
The Hamiltonian is given by
\begin{equation}
H=\epsilon_1 |1\rangle\langle1|\otimes I_2\otimes I_3 \otimes I_4+ I_1 \otimes I_2 \otimes I_3 \otimes \epsilon_4 |1\rangle\langle1|, \label{Hamiltonian_4}
\end{equation}
where $\epsilon_1, \epsilon_4>0$ and $I_1, I_2, I_3$ and $I_4$ are the identity operators. 
Under this Hamiltonian, the state $|\Psi(t)\rangle$ undergoes local unitary evolution for another time $\tau$ as given by
\begin{eqnarray}
&&|\Psi(t)\rangle_{1234}\rightarrow U_1\otimes I_2 \otimes I_3 \otimes U_4 |\Psi(t)\rangle_{1234} \nonumber \\
&&=e^{-i{\epsilon_1 \tau/\hbar |1\rangle\langle1|}}\otimes I_2 \otimes I_3 \otimes e^{-i{\epsilon_4 \tau/\hbar |1\rangle\langle1|}}|\Psi(t)\rangle_{1234}\nonumber\\
\label{evolution_4}
\end{eqnarray}
This state after evolution can be written as
\begin{eqnarray}
|\Psi^{'}\rangle_{1234}&=&\frac{-i \sin (\theta)}{2} \bigg \{e^{-i\epsilon_4 \tau/\hbar}|0011\rangle+e^{-i \epsilon_1 \tau/\hbar}|1100\rangle\bigg\}\nonumber\\
&&-\frac{\cos(\theta)}{2}\bigg\{(e^{-i(\epsilon_1+\epsilon_4) \tau/\hbar}|1001\rangle+|0110\rangle\bigg\} \nonumber\\
&&+\frac{e^{i\theta}}{2} \bigg\{e^{-i\epsilon_4 \tau/\hbar}|0101\rangle+e^{-i\epsilon_1 \tau /\hbar}|1010\rangle \bigg \},
\label{evolved_wavevectorsimple_ex4}
\end{eqnarray}
where $ \theta=2 J_m t$. The global phase shifts yield to the 
following expression
\begin{equation}
\Phi_T^{1234}=  \tan^{-1}
  \left[
    \frac{
      \splitfrac{
        -(1+\sin^2\theta)(\sin g_1+\sin g_4)
      }{
        - \sin(g_1+g_4)+\sin^2\theta\sin(g_1+g_4)
      }
    }{
    \splitfrac{
        (1+\sin^2\theta)(\cos g_1+\cos g_4)
      }{
        + \cos(g_1+g_4)-\sin^2\theta\cos(g_1+g_4)
      }
    }
  \right], 
\label{globalphase_ex4}
\end{equation}
and the local phase shifts are given by
\begin{eqnarray}
\Phi_T^1&=&tan^{-1}\bigg[\frac{-\sin g_1}{1+\cos g_1}\bigg],\nonumber\\
\Phi_T^4&=&tan^{-1}\bigg[\frac{-\sin g_4}{1+\cos g_4}\bigg],
\end{eqnarray}
where $ g_1=\epsilon_1 \tau/\hbar,~g_4= \epsilon_4 \tau/\hbar$.
Therefore, the Pancharatnam phase deficit is given by

\begin{eqnarray}
\Delta&=&\Phi_T^{1234}-[\Phi_T^1+\Phi_T^4]\nonumber\\
&=&  \tan^{-1}
  \left[
    \frac{
      \splitfrac{
        -(1+\sin^2\theta)(\sin g_1+\sin g_4)
      }{
        - \sin(g_1+g_4)+\sin^2\theta\sin(g_1+g_4)
      }
    }{
    \splitfrac{
        (1+\sin^2\theta)(\cos g_1+\cos g_4)
      }{
        + \cos(g_1+g_4)-\sin^2\theta\cos(g_1+g_4)
      }
}
  \right]\nonumber \\
&&-tan^{-1}\bigg[\frac{-\sin g_1}{1+\cos g_1}\bigg]-tan^{-1}\bigg[\frac{-\sin g_4}{1+\cos g_4}\bigg].\label{difference_ex4}
\end{eqnarray}
If we choose $g_1=g_4=\pi/2 $ in Eq. \eqref{difference_ex4}, 
it provides a connection between the Pancharatnam phase deficit and the concurrence as in Eq. \eqref{concurrence} given by
\begin{equation}
E=\frac{tan \Delta+2}{tan \Delta-2}.
\label{connection Delta_E}
\end{equation}

\begin{figure}\label{Plot_ex2}
\centering
\includegraphics[scale=0.9]{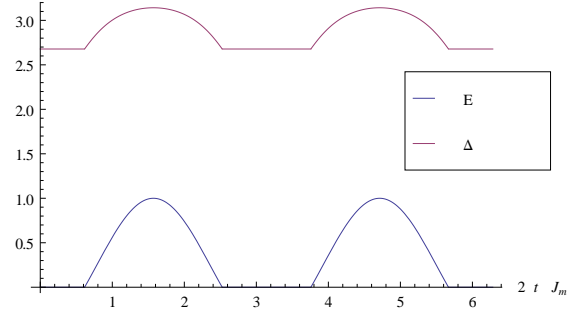}
\caption{(Color Online) Variation of the Pancharatnam phase deficit $\Delta$ and concurrence $E$ with $\theta=2J_mt$ for $g_1 = g_4 = \pi/2 $.
The blue line indicates the concurrence which is always nonzero and the magenta line indicates the value of Pancharatnam phase deficit.}
\end{figure}
As evident from this equation and also illustrated in Fig. 2, nonzero Pancharatnam phase deficit is a clear signature 
of macroscopic entanglement in this system.
\bigskip
\section{Conclusions}
In quantum information, detection of entanglement in bipartite and multipartite
states is very important. We have shown that for pure states a nonzero Pancharatnam phase deficit can detect entanglement.
It happens to be a sufficient condition for entanglement and 
we use this to get an insight of entanglement in macroscopic superpositions of quantum states. 
All the entangled systems may not show the phase deficit nonzero but if it holds a nonzero value, 
macroscopic entanglement is witnessed. This can be a very useful method to detect entanglement in macroscopic 
quantum systems. We have illustrated our method to detect entanglement for macroscopic superposition of coherent states 
and for macroscopically distant boundary spin states. For these specific cases, we have shown that the entanglement entropy 
and the concurrence can be expressed directly in terms of the Pancharatnam phase deficit. Since, the
Pancharatnam phase deficit can be measured in an 
experimental set up with the help of Mach-Zehnder interferometer,
this method could easily be applied to detect entanglement of macroscopic quantum systems in the quantum interferometry.

\begin{section}{Acknowledgment}
NS acknowledges the research fellowship of Department of Atomic Energy.

\end{section}

\bibliographystyle{h-physrev4}

\end{document}